\documentclass[3p,times,twocolumn]{elsarticle}

\usepackage{ecrc}

\volume{00}

\firstpage{1}

\journalname{Nuclear Physics B Proceedings Supplement}

\runauth{K. Kiers}

\jid{nuphbp}

\jnltitlelogo{Nuclear Physics B Proceedings Supplement}

\usepackage{amssymb}

\biboptions{comma,square}

\usepackage[figuresright]{rotating}

\begin{document}

\begin{frontmatter}



\dochead{}

\title{\boldmath CP violation in hadronic $\tau$
  decays\tnoteref{label1}} \tnotetext[label1]{Talk given at TAU2012,
  the 12th International Workshop on Tau Lepton Physics, Nagoya, Japan
  (September, 2012).  This talk is based on work done in collaboration
  with Alakabha Datta, Kevin Little, David London, Makiko Nagashima,
  Patrick J. O'Donnell and Alejandro
  Szynkman~\cite{datta,kiers_kpipi}.}

\author{Ken Kiers}
\ead{knkiers@taylor.edu}
\address{Physics and Engineering Department, Taylor University, 236 West Reade Ave., Upland, IN 46989, USA}

\begin{abstract}
We consider CP-violating effects in $\tau\to 3\pi\nu_\tau$ and
$4\pi\nu_\tau$ ($\Delta S = 0$), and $\tau\to K \pi\pi\nu_\tau$
($\Delta S = 1$), assuming that the usual Standard Model amplitudes
for these processes interfere with analogous amplitudes mediated by a
charged Higgs boson.  In the $\Delta S = 0$ case we focus specifically
on the intermediate resonant processes $\tau \to V\pi \nu_\tau$ (with
$V=\omega, \rho,$ and $a_1$), and consider three CP-odd observables --
the partial rate asymmetry, a polarization-dependent asymmetry and a
triple-product asymmetry.  In the $\Delta S = 1$ case we examine the
partial rate asymmetry, two ``modified'' rate asymmetries, and a
triple-product asymmetry.  The partial rate asymmetry is expected to
be small for both the $\Delta S = 0$ and $\Delta S = 1$ cases.
Evaluation of the other asymmetries indicates that they could
potentially be measurable.
\end{abstract}




\end{frontmatter}


\section{Introduction}
\label{intro}
It is currently expected that the Standard Model (SM) is a low-energy
approximation to a more fundamental theory.  Models of New Physics
(NP) typically predict the existence of new, heavy particles.  In
general, there are two approaches to searching for NP.  One approach
is to produce the new particle(s) directly at a collider.  Another
approach is to search for NP effects at low energies through the
virtual production of new particles.  Extensions of the SM typically
contain new sources of CP violation.  Thus, in the ``low-energy''
approach to searching for NP, searching for CP-odd signals can be
particularly effective~\cite{bigi_TAU}, particularly if the SM process
does not violate CP.

In general, non-zero CP-violating asymmetries require that more than
one amplitude contributes to the process in question.  Furthermore,
there needs to be a relative weak phase between contributing
amplitudes.  The simplest type of CP asymmetry is the partial rate
asymmetry (PRA), which is non-zero if the partial rates for the
process and CP-conjugate process are different.  The PRA is a
``global'' CP asymmetry (see Ref.~\cite{bigi_TAU}).  Alternatively, it
is possible to construct local asymmetries~\cite{bigi_TAU}, which can
in principle be larger.

\section{\boldmath CP violation in $\tau \to N\pi\nu_\tau~(\Delta S = 0)$}
\label{sec:DelS=0}

\begin{figure}[t]
\begin{center}
    \resizebox{3in}{!}{\includegraphics*{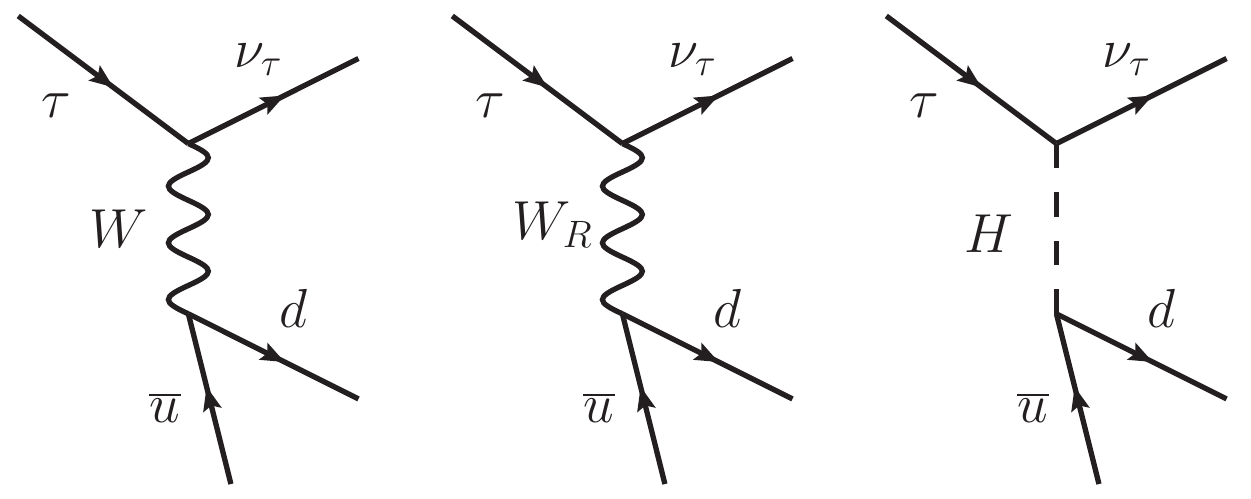}}
\caption{SM $W$-exchange diagram (left) and two possible NP
  contributions for the $\Delta S=0$ case.}
\label{fig:3diags}
\end{center}
\end{figure}

The $\Delta S =0$ hadronic $\tau$ decays are dominated by the
$W$-exchange diagram in the SM (see Fig.~\ref{fig:3diags}).  To the
extent that the decays are each described by a single amplitude within
the context of the SM, CP-violating effects require the interference
of the SM amplitude with a NP amplitude.\footnote{In this work we do
  not consider the SM CP-violating asymmetry discussed in
  Ref.~\cite{bigisanda}.}  Some possible NP contributions that one
could consider are those involving a new $W_R$-exchange diagram or a
charged Higgs diagram (see Fig.~\ref{fig:3diags}).  CP-violating
effects are likely to be very small in the $W_R$ case, since they are
suppressed by the neutrino mass or by $W_L-W_R$ mixing, so we will
consider only the case where the NP effects are due to the exchange of
a charged Higgs boson.  Higgs couplings to fermions would typically be
proportional to quark or lepton masses.  In this case, such couplings
would lead to a suppression, since $m_u$ and $m_d$ are so small.
Thus, in order for there to be a non-negligible effect, we will assume
that the exchanged Higgs has ``non-standard'' couplings.

%
%

Let us focus on the case of resonant production, concentrating in
particular on the modes $\tau\to \omega \pi\nu_\tau$, $\tau\to a_1
\pi\nu_\tau$ and $\tau\to \rho \pi\nu_\tau$.  For the purpose of the
numerical work, we will consider the $\omega$ and $a_1$ cases.  The
general structure for the SM hadronic current for $\tau\to
V\pi\nu_\tau$ (with $V=\omega, a_1, \rho$) is given by~\cite{decker,
  deckermirkes},
\begin{eqnarray}
    J^{\mu} & \equiv & \langle V(q_1)\pi(q_2) |\overline{d}\gamma^\mu
     (1-\gamma^5)u|0\rangle\nonumber\\
   & = & F_1(Q^2)\left(Q^2\epsilon_1^\mu-\epsilon_1\cdot q_2 Q^\mu\right) 
      \nonumber\\
     & &   \!\!\!\!\!\!\!\!\!\!+F_2(Q^2)\,\epsilon_1\cdot q_2
         \left(q_1^\mu-q_2^\mu-Q^\mu\frac{Q\cdot(q_1-q_2)}{Q^2}\right) \nonumber\\
     & & \!\!\!\!\!\!\!\!\!\!+iF_3(Q^2)\,\varepsilon^{\mu\alpha\beta\gamma}
             \epsilon_{1\alpha}q_{1\beta}q_{2\gamma}
       +F_4(Q^2)\,\epsilon_1\cdot q_2 Q^\mu \!,
\nonumber
\end{eqnarray}
where $Q\equiv q_1+q_2$, $F_1$-$F_4$ are form factors and $\epsilon_1$
denotes the polarization tensor for $V$.  Note that $F_4$ is expected
to be very small~\cite{deckermirkes}.  In order to take into account
the NP contribution, we parameterize the hadronic current for the
Higgs contribution as follows~\cite{datta},
\begin{eqnarray}
J_{\textrm{\scriptsize{Higgs}}} & = & \langle
V(q_1)\pi(q_2)|\overline{d}(a+b\gamma^5)u|0\rangle \nonumber\\
&=& 
\cases{
bf_H \, \epsilon_1 \cdot q_2, & $V$: vector, \cr
af_H \, \epsilon_1 \cdot q_2, & $V$: axial-vector,}\nonumber
\end{eqnarray}
where $a$ and $b$ parameterize the Higgs couplings and $f_H$ is a form
factor.  The parameters $a$ and $b$ can in principle contain
CP-violating (``weak'') phases, while the form factors $F_1$-$F_4$ and
$f_H$ can have CP-conserving (``strong'') phases.

\subsection{ Partial Rate Asymmetry}  

Let us consider first the PRA, which can be zero if the widths for the
process ($\Gamma$) and CP-conjugate process ($\overline{\Gamma}$) are
different.  The PRA is defined as follows,
\begin{eqnarray}
A_{CP} \equiv \frac{\Gamma-\overline{\Gamma}}{\Gamma+\overline{\Gamma}}
  = \frac{\Delta\Gamma}{\Gamma+\overline{\Gamma}}.
\label{eq:PRA}
\end{eqnarray}
For the vector meson case (e.g., $V=\omega$), we find,
\begin{eqnarray}
\Delta \Gamma \!= \!\!\!\int \!\!g(Q^2) \!\left|b f_H F_4\right|\!\sin\!\left(\delta_4-\delta_H\right)
  \sin\!\left(\phi_b\right) dQ^2,
\nonumber
\end{eqnarray}
where $\delta_4$ and $\delta_H$ represent possible strong phases
associated with $F_4$ and $f_H$, respectively, and where $\phi_b$
represents the weak phase associated with the parameter $b$.  (For
more details regarding the form of the function $g(Q^2)$, please see
Ref.~\cite{datta}.)  The above expression is suppressed by $F_4$, and
so the PRA is likely to be quite small.  A similar result holds for
the axial vector case.

\subsection{Polarization-dependent rate asymmetry}

The PRA is small because it involves the interference of terms
containing $f_H$ and $F_4$.  A more promising approach could thus be
to construct CP-violating quantities involving $f_H$ and $F_1$, $F_2$
or $F_3$.  One possibility is to weight the integral for the
differential width in such a way that such terms are extracted.  The
use of one particular angle allows us to construct an asymmetry that
depends on SM-NP cross-terms containing $F_1 f_H^*$ and $F_2f_H^*$.
Working in the hadronic rest frame, we weight the integral for the
differential width by $\cos(\beta)$, where $\beta$ is an angle defined
in the hadronic rest frame~\cite{datta}.  This of
course assumes that $\beta$ can be measured, which could be a
challenge if $V$ corresponds to a broad resonance.  Having weighted
the integral for the differential width in this manner, we can then
construct a CP-odd quantity by subracting the corresponding quantity
for the CP-conjugate process,
\begin{eqnarray}
    A_{CP}^{\langle\cos\beta\rangle} = 
      \frac{\Delta \Gamma_{\langle\cos\beta\rangle}}{\Gamma_{\textrm{\scriptsize{sum}}}} \; ,
      \label{eq:acpbeta}
\end{eqnarray}
where $\Gamma_{\textrm{\scriptsize{sum}}}\equiv
\Gamma+\overline{\Gamma}$.  For the numerator of the above expression,
we find,
\begin{eqnarray}
   \Delta\Gamma_{\langle\cos\beta\rangle} & = & \int \big[ g_1(Q^2) 
     |F_1|\sin(\delta_1-\delta_H) \nonumber\\
     & & \!\!\!\!\!\!\!\!\!\!\!\!\!\!\!\!\!\!\!\!
     + g_2(Q^2) |F_2|\sin(\delta_2-\delta_H)\big] |af_H|\sin(\phi_a)dQ^2,
   \nonumber
\end{eqnarray}
where the functions $g_1(Q^2)$ and $g_2(Q^2)$ depend on the
polarization of the $\tau$, $\delta_1$ and $\delta_2$ are strong
phases, and $\phi_a$ is a weak phase (see Ref.~\cite{datta} for
details).  The form factors $F_1$ and $F_2$ are expected to be
appreciable for the case $V=a_1$; for further discussion and a
numerical analysis, see Ref.~\cite{datta}, where it is noted that
asymmetries of order 15\% are consistent with experimental
uncertainties for the $V=a_1$ case.  An experimental challenge in this
case will be that the $a_1$ is a broad resonance.

\subsection{Triple-product asymmetry}

If the momentum of the $\tau$ can be measured, a triple-product
asymmetry can be constructed that depends on the $F_3$ term in the SM
hadronic current.  Here we consider the case $V=\omega$, since $F_3$
is the dominant form factor in this case.  Furthermore, the $\omega$
is a narrow resonance.  Assuming that the $\tau$ is unpolarized, we
find~\cite{datta},\footnote{In this and some subsequent expressions,
  the ``$\sim$'' symbol is used to denote the fact that various
  factors and the integration over phase space have been omitted for
  clarity.}
\begin{eqnarray}
  \Gamma_{\mbox{\scriptsize TP}} \sim \mbox{Im}(bf_H F_3^*)(\vec{\epsilon}_1\cdot\vec{q_1})
     \vec{\epsilon}_1\cdot (\vec{p}_\tau\times \vec{q}_1),
\end{eqnarray}
where $\vec{p}_\tau$ is the momentum of the $\tau$ in the hadronic
rest frame.  A CP-odd asymmetry can be formed by comparing the process
and CP-conjugate process,
\begin{eqnarray}
    A_{CP}^{\textrm{\scriptsize{TP}}} = 
      \frac{\Delta \Gamma_{\textrm{\scriptsize{TP}}}}{\Gamma_{\textrm{\scriptsize{sum}}}},
\end{eqnarray}
where,
\begin{eqnarray}
  \Delta \Gamma_{\textrm{\scriptsize{TP}}} \!\sim\! 
   (\vec{\epsilon}_1\!\cdot\!\vec{n}_1)(\vec{\epsilon}_1\!\cdot\!\vec{n}_2)
  |bf_HF_3|\!\cos(\delta_3\!-\!\delta_H) \!\sin(\phi_b),
  \nonumber
\end{eqnarray}
with $\vec{n}_1$ and $\vec{n}_2$ being direction vectors and
$\delta_3$ and $\phi_b$ being a strong and weak phase, respectively.
For this asymmetry one needs to measure the polarization of the
$\omega$ (see Ref.~\cite{wu} for a related discussion).  Numerical
estimates made in Ref.~\cite{datta} indicate that uncertainties in the
partial width allow for an asymmetry of order 30\% multiplied by
$(\vec{\epsilon}_1\cdot\vec{n}_1)(\vec{\epsilon}_1\cdot\vec{n}_2)$.

\section{\boldmath CP violation in $\tau\to K\pi\pi\nu_\tau~(\Delta S = 1)$}

Let us now consider CP violation in the case $\tau\to
K\pi\pi\nu_\tau$, which is similar to the $\Delta S = 0$ case
considered above, except that $d\to s$ in Fig.~\ref{fig:3diags}.  Once
again we will concentrate on the case in which the NP contribution is
due to the exchange of a charged Higgs boson.  For this case we will
consider the PRA, two ``modified'' rate asymmetries and a
triple-product asymmetry, all of which will be CP-odd.

The general structure of the SM hadronic current for $\tau^-\to
K^-\pi^-\pi^+\nu_\tau$ may be expressed in terms of form factors
$F_1$-$F_4$ as follows~\cite{kuhnmirkes1992},
\begin{eqnarray}
   J^\mu & \equiv & \langle K^-(p_1) \pi^-(p_2) \pi^+(p_3)| \bar{s}\gamma^\mu
     (1-\gamma^5) u | 0\rangle \nonumber \\
     &=& \big[F_1(s_1,s_2,Q^2) (p_1-p_3)_\nu \nonumber\\
     && + F_2(s_1,s_2,Q^2) (p_2-p_3)_\nu \big]T^{\mu\nu} \nonumber \\
     && + i F_3(s_1,s_2,Q^2) \epsilon^{\mu\nu\rho\sigma}
         p_{1\nu}p_{2\rho}p_{3\sigma} \nonumber\\
     && + F_4(s_1,s_2,Q^2) Q^\mu\, ,
\label{eq:jmu_sm}
\end{eqnarray}
where $Q^\mu=(p_1+p_2+p_3)^\mu$, $T^{\mu\nu}=g^{\mu\nu}-Q^\mu
Q^\nu/Q^2$ and $s_{1,2}=(Q-p_{1,2})^2$.  Once again, $F_4$ is expected
to be small; also, a triple-product can be formed using the $F_3$
term.  The form factors $F_1$-$F_3$ can arise via various resonances,
as described in Ref.~\cite{deckermirkessauerwas}.  The Higgs
contribution can be expressed through an effective Hamiltonian as
follows~\cite{kiers_kpipi},
\begin{eqnarray}
   {\cal H}^{\textrm{\scriptsize{NP}}}_{\textrm{\scriptsize{eff}}} \!=\!
    \frac{G_F}{\sqrt{2}}\!\sin\theta_c 
      \bar{\nu}_\tau(1\!+\!\gamma_5)\tau 
     \big[ \eta_S \bar{s}u  + \eta_P
     \bar{s} \gamma_5 u \big]\!+\!\mbox{h.c.,}
\nonumber
\end{eqnarray}
where the parameters $\eta_S$ and $\eta_P$ could in principle
contain weak phases.  Defining the pseudoscalar form factor,
\begin{eqnarray}
   \langle K^-(p_1) \pi^-(p_2) \pi^+(p_3)| \bar{s}
     \gamma^5 u | 0\rangle = f_H,
\end{eqnarray}
we see that CP-odd observables will involve cross-terms between the
(SM) form factors $F_1$-$F_4$ and the (NP) form factor $f_H$.

\subsection{Partial rate asymmetry}

Defining the PRA as above [see Eq.~(\ref{eq:PRA})], we find,
\begin{eqnarray}
  \Delta \Gamma \sim |F_4f_H\eta_P|\sin(\delta_4-\delta_H)\sin(\phi_P),
\end{eqnarray}
where $\delta_4$ and $\delta_H$ are strong phases associated with
$F_4$ and $f_H$, respectively, and $\phi_P$ is the weak phase
associated with $\eta_P$.  As was the case above, the PRA is
suppressed by $F_4$, and is thus likely to be very small.

\subsection{Three other CP-odd asymmetries}

Since the PRA is likely to be small, it may be more fruitful to use
asymmetric integration prescriptions for the integrals over phase
space in order to extract other SM-NP cross terms (i.e., those
depending on $F_{1-3}$ instead of $F_4$) from the differential width.
Reference~\cite{kiers_kpipi} defines three such prescriptions in terms
of two of the Euler angles in the hadronic rest frame (see also
Refs.~\cite{kuhnmirkes1992,kilian}).  These integration prescriptions
pick out different terms from the differential width.  We denote these
altered differential widths by $d\Gamma_i$, with $i=1,2,3$.  The
altered differential widths can then be compared to their counterparts
from the CP-conjugate process, and CP-odd asymmetries can be formed as
follows,
\begin{eqnarray}
   A_{CP}^{(i)} & = & \frac{1}{\Gamma +\overline{\Gamma}}
     \int \bigg(\frac{d\Gamma_i}{dQ^2\,ds_1 \,ds_2} \nonumber\\
     && ~~~~~~~~-\frac{d\overline{\Gamma}_i}{dQ^2\,ds_1 \,ds_2}\bigg)
      dQ^2\,ds_1 \,ds_2 \,.
\label{eq:ACP} 
\end{eqnarray}
Assuming, for simplicity, that $f_H$ is real and positive, and
omitting various factors, as well as the integration over phase space
(please see Ref.~\cite{kiers_kpipi} for further details), we find,
\begin{eqnarray}
  A_{CP}^{(1)} \!\!\!\!& \sim & \!\!\!\!\mbox{Im}(F_1-F_2) f_H \mbox{Im}(\eta_P) ,\nonumber\\
  A_{CP}^{(2)} \!\!\!\!& \sim & \!\!\!\!\mbox{Im}[F_1(p_1-p_3)^x\!+\!F_2(p_2-p_3)^x] f_H \mbox{Im}(\eta_P) ,\nonumber\\
  A_{CP}^{(3)} \!\!\!\!& \sim & \!\!\!\!\mbox{Re}(F_3) f_H \mbox{Im}(\eta_P) \nonumber,
\end{eqnarray}
where the various momenta are defined in the hadronic rest frame.  We
refer to the first two asymmetries as modified rate asymmetries.  The
third is a triple-product asymmetry.  All three asymmetries, being
CP-odd, depend on the imaginary part of the NP parameter $\eta_P$.

A numerical analysis of the three asymmetries was performed in
Ref.~\cite{kiers_kpipi}.  The analysis was guided by CLEO data for the
forms of $F_1$ and $F_2$~\cite{CLEO} and by theoretical
considerations~\cite{deckermirkessauerwas} for $F_3$.  The
normalization for the $F_3$ contribution was fixed such that it
contributed 5\% to the partial width.  Furthermore, $F_4$ was set to
zero and $f_H$ was assumed to be constant, for simplicity.  The
overall normalization was fixed by the BABAR result for the branching
ratio~\cite{BABAR}.  More recent BABAR~\cite{nugent, nugent_TAU} and
Belle~\cite{Belle} data was not used in the analysis.  Under the
assumption that $f_H$ is a real, positive constant, each of the three
asymmetries has the following form upon integration over phase space,
\begin{eqnarray}
  A_{CP}^{(i)} = a_{CP}^{(i)} f_H \mbox{Im}(\eta_P),
\end{eqnarray}
where $a_{CP}^{(i)}$ is a numerical factor obtained by integrating
over phase space.  As noted in Ref.~\cite{kiers_kpipi}, it is useful
to define differential quantities, $da_{CP}^{(i)}/dX$, with
$X=M_{K\pi\pi}, M_{K\pi}, M_{\pi\pi}$ and $\cos\theta$ (where $\theta$
is an angle defined in the $\tau$ rest frame~\cite{kiers_kpipi}).
Figure 3 in Ref.~\cite{kiers_kpipi} shows plots of the differential
quantities $da_{CP}^{(i)}/dX$.  It is evident from these plots that
each of the coefficients $a_{CP}^{(i)}$ suffers some amount of cancellation upon
integration over phase space.  Reference~\cite{kiers_kpipi} contains a
discussion of how one might mitigate these cancellations to some
degree by altering the definitions of the asymmetries.  Assuming that
$|f_H\eta_P|$ is such that the NP contribution to the BR saturates the
experimental uncertainty, it was found that the maximum values for the
(altered) CP-odd asymmetries ranged from being of order 1\% to being
of order 5\%.  More realistic values for $f_H$ and $\eta_P$ would
likely tend to reduce the maximal possible asymmetries somewhat
further.

\section{Concluding remarks}

Hadronic $\tau$ decays provide a promising avenue to search for CP-odd
signals of NP, particularly for evidence of a charged Higgs.  While
partial rate asymmetries are likely to be small in the decays modes
considered here, one can construct other CP-odd asymmetries as well,
and in these cases, moderate to significant signals are possible.
Further refinement of the decay distributions and determination of
form factors will allow for a more precise analysis of CP-odd
quantities.



\bigskip
\noindent
{\bf Acknowledgments:}
I would like to thank the conference organizers for their invitation
and support.  I am indebted to my collaborators -- K. Little,
A. Datta, D. London, M. Nagashima, P.J. O'Donnell and A. Szynkman.  I
would also like to thank the following people for helpful
correspondence and/or discussion: H. Hayashii, M. Roney, S. Banerjee,
I. Nugent and I. I. Bigi.  The work described here was supported by the
U.S.\ National Science Foundation under Grants PHY-0301964,
PHY--0601103 and PHY-1215785.



\bibliographystyle{elsarticle-num}




\end{document}